\documentclass{mn2e}
\usepackage{psfig}
\input{epsf}

\title[Higher Order Clustering in the 2dFGRS]
{The 2dF Galaxy Redshift Survey: Higher order galaxy correlation functions}

\author[Croton et al.]{
\parbox[t]{\textwidth}{
D. J. Croton$^{1}$,
E. Gazta\~{n}aga$^{2,3}$,
C. M. Baugh$^4$,
P. Norberg$^{5}$,
M. Colless$^6$,
I. K.\ Baldry$^7$,
J. Bland-Hawthorn$^8$,
T. Bridges$^9$, 
R. Cannon$^8$, 
S. Cole$^4$, 
C. Collins$^{10}$, 
W. Couch$^{11}$, 
G. Dalton$^{12,13}$,
R. De Propris$^6$,
S. P.\ Driver$^6$, 
G. Efstathiou$^{15}$, 
R. S.\ Ellis$^{17}$, 
C. S.\ Frenk$^4$, 
K. Glazebrook$^7$, 
C. Jackson$^{14}$,
O. Lahav$^{15,16}$, 
I. Lewis$^{12}$, 
S. Lumsden$^{18}$, 
S. Maddox$^{19}$,
D. Madgwick$^{20}$,
J. A.\ Peacock$^{21}$,
B. A.\ Peterson$^6$, 
W. Sutherland$^{21}$,
K. Taylor$^{17}$.
(The 2dFGRS Team)
}
\vspace*{6pt} \\ 
$^1$Max-Planck-Institut f\"ur Astrophysik, D-85740 Garching, Germany \\
$^2$INAOE, Astrofisica, Tonantzintla, Apdo Postal 216 y 51, Puebla 7200,
    Mexico \\
$^3$Institut d'Estudis Espacials de Catalunya, ICE/CSIC, Edf.
    Nexus-104-c/Gran Capita 2-4, 08034 Barcelona, Spain  \\
$^4$Department of Physics, University of Durham, South Road, 
    Durham DH1 3LE, UK \\ 
$^5$ETHZ Institut f\"ur Astronomie, HPF G3.1, ETH H\"onggerberg, CH-8093
       Z\"urich, Switzerland \\
$^6$Research School of Astronomy \& Astrophysics, The Australian 
    National University, Weston Creek, ACT 2611, Australia \\
$^7$Department of Physics \& Astronomy, Johns Hopkins University,
       Baltimore, MD 21118-2686, USA \\
$^8$Anglo-Australian Observatory, P.O.\ Box 296, Epping, NSW 2111,
    Australia\\  
$^{9}$Department of Physics, Queen's University, Kingston, 
    Ontario K7L 3N6, Canada \\
$^{10}$Astrophysics Research Institute, Liverpool John Moores University,  
    Twelve Quays House, Birkenhead, L14 1LD, UK \\
$^{11}$Department of Astrophysics, University of New South Wales, Sydney, 
    NSW 2052, Australia \\
$^{12}$Department of Physics, University of Oxford, Keble Road, 
    Oxford OX1 3RH, UK \\
$^{13}$Space Science \& Technology Division, Rutherford Appleton Laboratory, 
    Chilton OX11 0QX, UK \\
$^{14}$CSIRO Australia Telescope National Facility, PO
    Box 76, Epping, NSW 1710, Australia \\
$^{15}$Institute of Astronomy, University of Cambridge, Madingley Road,
    Cambridge CB3 0HA, UK \\
$^{16}$Department of Physics and Astronomy, University College London, 
    Gower Street, London WC1E 6BT, UK \\
$^{17}$Department of Astronomy, California Institute of Technology, 
    Pasadena, CA 91025, USA \\
$^{18}$Department of Physics, University of Leeds, Woodhouse Lane,
       Leeds, LS2 9JT, UK \\
$^{19}$School of Physics \& Astronomy, University of Nottingham,
       Nottingham NG7 2RD, UK \\
$^{20}$Department of Astronomy, University of California, Berkeley, 
       CA 94720, USA \\
$^{21}$Institute for Astronomy, University of Edinburgh, Royal Observatory, 
       Blackford Hill, Edinburgh EH9 3HJ, UK.
\vspace{-0.5cm}
}

\date{Accepted ---. Received ---;in original form ---}

\pubyear{2003}

\newcommand{\xibar}{\bar{\xi}}

\newcommand{\plotone}[1]
           {\centering \leavevmode \psfig{file=#1,width=\columnwidth,clip=}}

\newcommand{\plotfull}[1]
           {\centering \leavevmode \psfig{file=#1,width=\textwidth,clip=}}

\begin{document}

\maketitle

\begin{abstract}
We measure moments of the galaxy count probability distribution
function in the two-degree field galaxy redshift survey (2dFGRS).  The
survey is divided into volume limited subsamples in order to examine
the dependence of the higher order clustering on galaxy luminosity.
We demonstrate the hierarchical scaling of the averaged $p$-point galaxy
correlation functions, $\xibar_{p}$,  up to $p=6$.  The hierarchical amplitudes,
\smash{$S_p = \xibar_{p}/\xibar_2^{p-1}$}, are approximately
independent of the cell radius used to smooth the galaxy distribution
on small to medium scales. On larger scales we find the higher order
moments can be strongly affected by the presence of rare, massive
superstructures in the galaxy distribution.  The skewness $S_3$ has a weak
dependence on luminosity, approximated by a linear dependence on log
luminosity.  
We discuss the implications of our results for simple models of linear and 
non-linear bias that relate the galaxy distribution to the underlying mass.

\end{abstract}


\section{Introduction}

The pattern of galaxy clustering can be quantified by measuring the
galaxy count probability distribution function (CPDF) on a range of
smoothing scales.  The CPDF gives the probability that a randomly
chosen region of the universe will contain a particular number of
galaxies, and is typically expressed as a function of both the size of
the region smoothed over and the galaxy number within that
volume.  Traditionally, most effort has been directed at measuring the
second moment of the count distribution, the variance, $\xibar_2$,
through the autocorrelation function or, equivalently, its Fourier
transform, the power spectrum (e.g. Percival et al. 2001; 
Padilla \& Baugh 2003; Tegmark et al. 2004). 
The higher order moments of the CPDF,
expressed as volume averaged correlation functions, $\xibar_p$
($p=2,3, \dots$), provide a much more detailed description of galaxy
clustering, probing the shape of the low and high count tails of the
distribution.

The higher order moments of the dark matter distribution are known to
display a hierarchical scaling in which the $p$-point volume averaged
correlation functions, $\xibar_p$, can be written in terms of the
variance of the count distribution, $\xibar_2$: $\xibar_p = S_p
\xibar^{p-1}_2$ (e.g. see Peebles 1980, Juszkiewicz, Bouchet \& Colombi 1993,
Bernardeau 1994, Baugh, Gazta\~naga \& Efstathiou 1995, Gazta\~naga \&
Baugh 1995, Fosalba \& Gazta\~naga 1998).  
This scaling is a signature of the evolution under
gravitational instability of an initially Gaussian distribution of
density fluctuations.  A remarkable feature of the scaling is that the
values of the hierarchical amplitudes, $S_p$, on scales for which the
density field evolves linearly or in a quasi-linear fashion, are
insensitive to cosmic epoch and essentially independent of the
cosmological density parameter or the value of the cosmological
constant. For a comprehensive review of such results see 
Bernardeau et al. (2002) and references therein.

Departures from the hierarchical scaling of the higher order moments
could conceivably arise in three ways: 

\begin{itemize}

\item[(i)] A strongly non-Gaussian distribution of primordial density 
waves as could arise, for instance, due to a seed non-linear 
fluctuation such as a global texture (see Gazta\~{n}aga \& M{a}h{o}nen 
1996, Gazta\~naga \& Fosalba 1998,
Scoccimarro, Sefusatti \& Zaldarriaga 2003 for examples 
of how the $S_p$ scale in this case).  
This avenue now seems unlikely, following the clear detection
of multiple acoustic peaks in the power spectrum of cosmic microwave
background temperature fluctuations (Netterfield et al. 2002; Hinshaw
et al. 2003; Mason et al. 2003; Scott et al. 2003; Kuo et al. 2004);
such peaks are difficult to reconcile with models that include
cosmological defects (Kamionkowski \& Kowsowsky 1999). Moreover
strongly non-Gaussian primordial fluctuations are ruled out by the
first year WMAP results (Komatsu et al. 2003; Gazta\~naga \& Wagg 2003)

\item[(ii)] A weakly non-Gaussian distributed primordial density field, 
resulting from a non-linear perturbation to a Gaussian density field.
This scenario is difficult to distinguish from the evolution of an 
initially Gaussian field under gravitational instability, because the 
perturbation can introduce a shift to the amplitudes $S_p$ that is also 
hierarchical. This can happen even in the case where the non-linear 
perturbation produces a negligible effect on the power spectrum 
(Bernardeau et al. 2002).

\item[(iii)] The spatial bias between the galaxy distribution 
and the underlying distribution of dark matter. 
Fry \& Gazta\~{n}aga (1993) demonstrated that, under a local 
biasing prescription, the hierarchical scaling of the higher order 
moments is preserved but the amplitudes $S_p$ can change as a function of 
time or luminosity. This conclusion is also reached using more 
sophisticated, physically motivated semi-analytic models of 
galaxy formation (Kauffmann et al. 1999; Benson et al. 2000; 
Scoccimarro et al. 2001).

\end{itemize}

Previous attempts to measure the higher order correlation functions
have been hamstrung by the small size of the available redshift
surveys, a shortcoming that is exacerbated once volume limited
subsamples are constructed (Hui \& Gazta\~naga 1999).  Nevertheless,
early counts-in-cells studies established that the first few higher 
order moments of the galaxy distribution 
displayed the hierarchical scaling expected in
the gravitational instability framework (Groth \& Peebles 1977;
Peebles 1980; Gazta\~{n}aga 1992; 
Bouchet et al. 1993; Fry \& Gazta\~{n}aga 1994, Ghigna et
al. 1996, Feldman et al. 2001).  
Such analyses were typically limited to measuring the
three and four point correlation functions.  The nature of the
dependence of the hierarchical amplitudes on luminosity has not been
convincingly established.  Recent work to investigate this in the
optical (Hoyle et al. 2000) and in the far infrared (Szapudi et
al. 2000) was restricted to probing fairly narrow ranges of luminosity 
due to the size of the redshift surveys then available. 
 
The advent of multi-fibre spectrographs exploited by sustained
observing campaigns has led to a new generation of redshift survey
which represents order of magnitude advances over surveys completed in
the last millennium.  The Sloan Digital Sky Survey (York et al. 2000)
and the Two-degree Field Galaxy Redshift Survey (2dFGRS, Colless et
al. 2001) have provided maps of the clustering pattern of galaxies
with unprecedented detail.  Analysis of the 2dFGRS clustering has
suggested that the flux limited sample could be an essentially
unbiased tracer of the dark matter in the Universe (Lahav et al. 2002;
Verde et al. 2002)
\footnote{Note that with the weighting scheme 
adopted to compensate for the radial selection function, the 
characteristic luminosity of the flux limited 2dFGRS used in these 
studies is $\approx 2L_{*}$.}.  
These results confirmed previous deductions about galaxy bias
(e.g. Gazta\~naga 1994, Frieman \& Gazta\~naga 1999, Gazta\~naga \&
Juszkiewicz 2001) reached using the parent angular catalogue of the
2dFGRS, the APM Galaxy Survey (Maddox et al. 1990, 1996).  The 2dFGRS
covers a volume that is an appreciable fraction of that sampled by the
APM Survey, with full redshift coverage (modulo the relatively small
redshift incompleteness that still remains).  This means that for the
first time, a measurement of the higher order moments is possible in
three dimensions with comparable accuracy to that attainable in two
dimensions, but without the added complication of the effects of
projection (Bernardeau \& Gazta\~naga 1996; Szapudi \& Gazta\~naga
1998).

The sheer number of galaxies in the 2dFGRS allows it to be subdivided
in order to probe the dependence of the clustering signal on intrinsic
galaxy properties in more detail.  Norberg et al. (2001) found that
the amplitude of the projected two point correlation function scales
with luminosity, and characterised this trend using a relative bias
factor with a linear dependence on luminosity.  In this paper we
extend the work of Norberg et al. to study the higher order
clustering of galaxies in the 2dFGRS and its dependence on luminosity.
Our approach is the same as that followed in Baugh et al. (2004), who 
measured the higher order correlation functions of a sample of $L_*$ 
galaxies and found that they follow a hierarchical scaling. 

We provide a brief review of the measurement of the moments of the
CPDF in Section 2. In Section 3, we discuss the specific application
of this method to the 2dFGRS; an important feature of our analysis is
the use of mock catalogues to estimate the errors on our measurements
(see Section 3.3). Our results for the higher order correlation
functions and the hierarchical amplitudes are given in Section 4. We
quantify the variation of the higher order moments with luminosity in
Section 5, and discuss the interpretation of these results in terms of
a simple relative bias model. Our conclusions are set out in Section
6.  Throughout, we adopt standard present day values of the cosmological
parameters to compute comoving distance from redshift: a density
parameter $\Omega_{m}=0.3$ and a cosmological constant
$\Omega_{\Lambda}=0.7$.

\section{Count-in-Cells Statistics}

The count probability distribution function (CPDF) and its moments
have been used extensively to quantify the clustering pattern of
galaxies (e.g. White 1979; Peebles 1980). In this Section we give an
outline of the counts-in-cells approach, explaining how the volume
averaged $p$-point correlations are derived from the CPDF and give a
brief theoretical background.  A more comprehensive discussion of the
counts-in-cells approach can be found in Bernardeau et al. (2002).

\subsection{Estimating the $p$-point volume averaged correlation functions} 

The $p$-point moment, or (un-reduced) correlation function,
\smash{$m_3(r_1,r_2,r_3) \equiv <\delta(r_1)...\delta(r_p)>$}, can be
used to fully characterise the clustering of a fluctuating field
$\delta(r)$.  The reduced $p$-point correlation function,
$\xi_p(r_1,...,r_p)$, is defined as the connected part of the above
$p$-point correlation in such a way that for $p>2$: $\xi_p=0$  for a Gaussian
field (see Bernardeau et al. 2002 for more details).  Following the
standard convention, for the remainder of this paper when we talk
about correlations we will always assume they are ``reduced" 
correlations.

The $p$-point \emph{volume averaged} galaxy correlation
function, $\xibar_{p}(V)$, can be written as the integral of the
$p$-point correlation function, $\xi_{p}$, over the sampling volume,
$V$ (Peebles 1980):

\begin{equation}
\xibar_{p}(V) = \frac{1}{V^{p}} \int_V {\rm d}^3r_1 \ \dots \ {\rm d}^3r_{p} \  \xi_p
(\mathbf{r_1},\ \dots\ ,\mathbf{r_{p}})~.
\label{volxi}
\end{equation}
A practical way in which to estimate
$\xibar_{p}(V)$ is to randomly throw cells down within the galaxy distribution,
recording the number of times a cell contains $N$ galaxies so as to
build up the galaxy CPDF, $P_{N}(V)$. Since we adopt spherical cells,
the CPDF is a function of the sphere radius, $R$,
\begin{equation}
P_N(R) = \frac{N_N}{N_T}~,
\end{equation}
where $N_{N}$ is the number of cells that contain $N$ galaxies out of
a total number of cells thrown down, $N_{T}$.  The volume averaged
correlation functions $\xibar_{p}(V)$ are then related to the moments
of the CPDF, $m_{p}$:
\begin{equation}
m_p(R) = \langle (N- \bar{N})^p \rangle = \sum_{N=0}^{\infty} P_N (R)
(N-\bar{N})^p~,
\end{equation}
where $\bar{N}$ is the mean number of galaxies in a cell of volume $V$
and is calculated directly from the CPDF
\begin{equation}
\bar{N}= \sum_{N=0}^{\infty} N P_N~.  
\end{equation}
For the case of a continuous distribution, $\xibar_p$ is related to
the corresponding cumulant, $\mu_{p}$, through $\bar{N}^p \xibar_p =
\mu_p$, where the cumulants are defined as (see Gazta\~{n}aga 1994 for
details):
\[ \mu_2 = m_2~  ~~~;~~~ \mu_3 = m_3~, \]
\begin{equation} 
 \mu_4 = m_4 - 3m_2^2~  ~~~;~~~
\mu_5 = m_5 - 10m_3 m_2~. 
\end{equation}
If instead we are dealing with a discrete distribution, these
relations must be corrected. A Poisson shot noise model is adopted
(see Baugh et al. 1995 for a discussion of this point), to give corrected 
estimates of the moments, $k_p$:
\[ k_2 = \mu_2 - \bar{N}  ~~~;~~~  k_3 = \mu_3 - 3k_2 - \bar{N}~, \]
\[ k_4 = \mu_4 - 7k_2 - 6k_3 - \bar{N}~, \]
\begin{equation} k_5 = \mu_5 - 15k_2 - 25k_3 - 10k_4 - \bar{N}~. \end{equation}
The volume-averaged correlation functions, calculated from the galaxy
CPDF, follow directly from the relation $\xibar_p = k_p/\bar{N}^p$.

\subsection{Scaling of the higher order moments}

In the hierarchical model of clustering, all higher-order correlations
can be expressed in terms of the 2-point function, $\xibar_2$, and
dimensionless scaling coefficients, $S_{p}$:
\begin{equation}
\xibar_{p} = S_{p} ~\xibar_2^{p-1}.
\label{eq:sp}
\end{equation}
Traditionally, $S_3 = \xibar_3 / \xibar_2^2$ is referred to as the
\emph{skewness} of the distribution and $S_4 = \xibar_4 / \xibar_2^3$
as the \emph{kurtosis}.  The hierarchical scaling of the higher order
moments arises from the evolution due to gravitational
instability of an initially Gaussian distribution of density
fluctuations (see Bernardeau et al. 2002 and references therein).

\subsection{Systematic effects: biased estimators}

In addition to sampling errors (see Section 3.3 below), the estimation of the
hierarchical amplitudes can be compromised by systematic effects, as
discussed in some detail by Hui \& Gazta\~{n}aga (1999). These authors
identified two sources of error that could lead to a systematic bias
in the inferred values of $S_p$. The first effect arises from
biases in the estimates of the higher order correlation functions themselves, known as
the ``integral constraint bias'' (see e.g. Bernstein 1994).  The
second effect originates in the nonlinear combination of
$\bar{\xi}_{p}$ and $\bar{\xi}_{2}$ to form $S_p$; this is called 
 the ``ratio bias''.  The latter effect
dominates on large scales and tends to cause the inferred values of
the $S_p$ to be biased low.  Hui \& Gazta\~{n}aga wrote down
expressions for these biases which
accurately reproduce the systematic effects seen upon estimating the
hierarchical amplitudes from sub-volumes extracted from N-body simulations.

As mentioned above, we will use different volume limited samples to
study the luminosity dependence of the hierarchical amplitudes, $S_p$.
As the luminosity that defines a sample is made brighter, the volume
of the sample increases. Thus the estimation biases 
tend to cause the $S_p$ to
increase with sample luminosity.  This spurious tendency has already
been reported in the literature (see Hui \& Gazta\~naga 1999). For
volumes of the size used in our analysis, it turns out that the
predicted biases are smaller than the corresponding sampling errors
(e.g. see figure~3 in Hui \& Gazta\~naga 1999).  This is the first time
that a redshift survey has been available which is large enough to
overcome such systematic biases.

\subsection{Galaxy biasing}

Galaxy samples constructed using different selection criteria display
different clustering patterns. This leads one to the conclusion that
distinct samples of galaxies must trace the underlying mass
distribution in different ways, a phenomenon that is generally known
as galaxy bias.

A simple, heuristic scheme describing the impact of a local bias on
the scaling of the higher order moments was proposed by Fry \&
Gazta\~{n}aga (1993). These authors demonstrated that in this case,
the scaling of the higher order moments of the galaxy distribution
should mirror that of the dark matter, though possibly with different
values for the hierarchical amplitudes $S_p$.  Fry \& Gazta\~naga made
the assumption that the density contrast in the galaxy distribution,
$\delta^{\rm G}$, i.e. the fractional fluctuation around the mean
density, could be written as a Taylor expansion of the density
contrast of the dark matter,
$\delta^{\rm DM}$: 
\begin{equation}
\delta^{\rm G}  = \sum_{k=0}^{\infty} \frac{b_{k}}{k!} (\delta^{\rm DM})^{k}.
\end{equation}
On scales where the variance, $\xibar_2^{\rm DM}$, is small, the
leading order contribution to the two-point volume averaged
correlation function of galaxies has the form:

\begin{equation}
\xibar^{\rm G}_2  = b_{1}^{2} ~\xibar^{\rm DM}_{2}, 
\label{eq:b1}
\end{equation} 
where $b_1$ is the ubiquitous linear bias $b$.  The leading order
forms for the hierarchical amplitudes, $S_{p}$, for the cases $p=3$
and $p=4$ are:

\begin{eqnarray}
S^{\rm G}_{3} &=& \frac{1}{b_{1}} \left( S^{\rm DM}_{3} + 3 c_{2}\right) \\ 
S^{\rm G}_{4} &=& \frac{1}{b_{1}^{2}} 
\left( 
S^{\rm DM}_{4} 
+ 12 c_{2} S^{\rm DM}_{3} 
+ 4 c_{3} + 
12 c^{2}_{2}
\right), 
\label{eqnfg93}
\end{eqnarray} 
where we use the notation $c_{k} = b_{k}/b_{1}$.  Expressions for the
hierarchical amplitudes are given up to $p=7$ in Fry \& Gazta\~{n}aga
(1993).

Mo, Jing \& White (1997) give theoretical predictions for the
coefficients $b_k$ using the Press \& Schechter (1974) formalism and 
exploiting the framework developed by Cole \& Kaiser (1989) and 
Mo \& White (1996).  For
halos of mass $M$, the first two bias factors ($k=1$ and 2) are given
by:
\begin{eqnarray}
\label{b1}
&& b_1  = 1 + {\nu^2 - 1 \over \delta_c} \\
\label{b2}
&& b_2  = 2 \left(1 - {17\over 21} \right) {\nu^2 - 1 \over \delta_c}
+ {\nu^2 \over \delta_c^2} (\nu^2 - 3)
\end{eqnarray}
where $\nu \equiv \delta_c / \sigma(M)$, $\delta_c$ is the
linear theory overdensity at the time of collapse ($\delta_c = 1.686$
for $\Omega=1$) and $\sigma (M)$ is the linear {\it rms} fluctuation
on the mass scale of the halos.  This is a simple model but nevertheless
it shows some tendencies that are correct.  For example, a
typical mass halo corresponding to $\nu=1$ displays an unbiased
variance with $b_1=1$, but introduces a bias in the skewness, since
$c_2 = b_2 = -0.7$.  As a further illustration, consider massive halos
defined by $\nu^2>3$; in this case the Mo, Jing \& White theory
predicts that $c_2>0$, while less massive halos could produce $c_2<0$.
To get more realistic values of $b_k$ for galaxy bias, a prescription 
has to be adopted for populating dark matter halos of a given mass 
with galaxies of a given luminosity (Benson et al.
2001; Scoccimarro et al. 2001; Berlind et al. 2003).

In the interpretation of the higher order moments presented in this
paper we will make use of a {\emph{relative}} bias, which describes
the change in clustering compared with that measured for a reference
sample (Norberg et al. 2001, 2002a).  Using Eq.~\ref{eq:b1} as a
guide, we define the relative bias, $b_r=b_1/b^*_1$, of a sample as
the square root of the ratio of the $2$-point correlation function
measured for the sample over that measured for the reference sample,
denoted by an asterisk (the reference sample will be defined
explicitly in Section 5):

\begin{equation}
b_r \equiv {b_1\over{b^*_1}} 
= \Big( \frac{\xibar^{G}_{2}}{\xibar^{G *}_{2} } \Big) ^{1/2}.
\label{eq:br}
\end{equation}
Thus, we can obtain an estimate of the relative bias from the ratio of
the variances.

When the linear bias is a good approximation ($c_k \simeq 0$ for
$k>1$), we can relate $S_p^G$ in different galaxy samples regardless
of the underlying DM value of $S_p$:
\begin{equation} 
S^{G}_{p} = \frac{S^{G *}_{p}}{b^{p-2}_{r}}.
\label{eq:SJlinear}
\end{equation} 
More generally, one can manipulate Eq.~10 to write down an expression
comparing $S_p^G$ for two galaxy samples, eliminating $S_3^{\rm DM}$
for the underlying dark matter (e.g. see Eq.~9 in Fry \& Gazta\~naga
1993).  For the skewness:
\begin{equation} 
S^{G * }_{3} = b_r S^{G}_{3} - 3 \frac{(c_{2} - c^{*}_{2})}{b_1^*} ~, 
\end{equation} 
where an asterisk denotes a quantity describing the reference sample,
and $b_{r} = b_{1}/b^{*}_{1}$ is the relative bias defined above.  Any
second order relative bias effects are thus given by:
\begin{equation} 
c_{2}' = \frac{(c_{2}- c^{*}_{2})}{b_1^*} = \frac{1}{3} \left(b_{r} S^{G}_{3} - 
S^{G *}_{3} \right).
\label{eq:c2'}
\end{equation} 
As a special case, if the reference sample is un-biased (i.e. $b_1^*=1$ and
$c_p^*=0$), we then have $c_{2}' =c_{2}$.

\begin{table*}
  \centering
  \footnotesize
  \caption[]{Properties of the combined 2dFGRS SGP and NGP volume-limited
  catalogues (VLCs).  Column 1 gives the numerical label of the
  sample. Columns 2 and 3 give the faint and bright absolute magnitude
  limits that define the sample. 
  The fourth column gives the median luminosity of each volume limited
  sample in units of $L_{*}$, computed using the Schechter function parameters
  quoted by Norberg et al. (2002b). 
  Columns 5, 6 and 7 give the
  number of galaxies, the mean number density and the mean
  inter-galaxy separation for each VLC, respectively.  Columns 8 and 9
  state the redshift boundaries of each sample for the nominal
  apparent magnitude limits of the survey; columns 10 and 11 give the
  corresponding comoving distances. Finally, column 12 gives the
  combined SGP and NGP volume.  All distances are comoving and are
  calculated assuming standard cosmological parameters
  ($\Omega_{m}=0.3$ and $\Omega_{\Lambda}=0.7$).}
  \begin{tabular}{ccccccccccccc} 
    \hline \hline
    VLC & 
    \multicolumn{2}{c}{Mag. range} & Median lum. &   
           {N$_{\rm G}$} &
           {$\rho_{\rm ave}$} & {d$_{\rm mean}$} & {z$_{\rm min}$} & {z$_{\rm max}$} &
           {D$_{\rm min}$} & {D$_{\rm max}$} & {Volume}\\
     ID  & 
    \multicolumn{2}{c}{\tiny $M_{b_{\rm J}}-5\log_{10}h$} & 
            $L/L_{*}$ & & 
           {\tiny $10^{-3} h^3$Mpc$^{-3}$} & {\tiny $h^{-1}$Mpc} & & &
           {\tiny $h^{-1}$Mpc} & {\tiny $h^{-1}$Mpc} & 
           {\tiny $10^6h^{-3}$Mpc$^3$}\\
           \hline
    1 &  $-$17.0 & $-$18.0 & 0.13 & 8038 & 10.9 & 4.51 & 0.009 & 0.058 & 24.8 & 169.9 & 0.74\\
    2 &  $-$18.0 & $-$19.0 & 0.33 & 23290 & 9.26 & 4.76 & 0.014 & 0.088 & 39.0 & 255.6 & 2.52\\
    3 &  $-$19.0 & $-$20.0 & 0.78 & 44931 & 5.64 & 5.62 & 0.021 & 0.130 & 61.1 & 375.6 & 7.97\\
    4 &  $-$20.0 & $-$21.0 & 1.78 & 33997 & 1.46 & 8.82 & 0.033 & 0.188 & 95.1 & 537.2 & 23.3\\
    5 &  $-$21.0 & $-$22.0 & 3.98 &  6895 & 0.110 & 20.9 & 0.050  & 0.266 & 146.4 & 747.9 & 62.8\\
   \hline \hline
  \end{tabular}
\end{table*}  

\begin{table*}
  \centering
  \footnotesize
  \caption[]{The best fit values  and 2$-\sigma$ error ($\Delta \chi^2 = 4$) 
for $S_p$ (columns 4 to 7).  The range of scales used in the fits is
given in columns 2 and 3.  The number in brackets after each error
gives the reduced $\chi^2$ value for the fit, using the number of
degrees of freedom derived from the principal component analysis.  The
last two columns give the relative linear bias, $b_r$ (defined by
Eq.~\ref{eq:br}) and the second order bias term, $c_{2}'$ (defined by
Eq.~17).  The reference sample is sample number 3.  These values are
obtained for the full volume limited samples.  A blank entry indicates
that a reliable measurement of the particular hierarchical amplitude
was not possible for the sample in question.  }
  \begin{tabular}{cccccccccc} 
    \hline \hline
    {VLC} & $R_{\rm min}$ & $R_{\rm max}$ &{$S_3$} & {$S_4$}& {$S_5$} &  ${S_6}$ & {$b_r$} & {$c_{2}'$}  \\
    { ID}    & {\tiny $h^{-1}$Mpc} & {\tiny $h^{-1}$Mpc} & &         &        &         &         &          \\
    \hline
     1 & 0.71 & 7.1 & $2.58\pm0.37 \,(0.1) $ & $9.3\pm4.0 \,(0.1) $ & $34\pm32 \,(0.1)$ &  $---$      &  $0.96\pm0.16 \,(0.1)$ & $0.17\pm0.25 \,(0.1)$\\
     2 & 0.71 & 7.1 & $2.38\pm0.25 \,(0.1) $ & $8.2\pm2.3 \,(0.9) $ & $36\pm20 \,(0.4)$ & $185\pm170 (0.1)$ &  $0.96\pm0.08 \,(0.3)$ & $0.11\pm0.13 \,(0.1)$\\
     3 & 0.71 & 7.1 & $1.95\pm0.18 \,(6.1) $ & $5.5\pm1.4 \,(2.3) $ & $18\pm11 \,(1.9)$ & $46\pm50 (1.1)$   &  $1$           &  $0$         \\
     4 & 0.80 & 8.9 & $2.01\pm0.17 \,(1.2) $ & $6.0\pm1.5 \,(0.6) $ & $22\pm12 \,(0.4)$ &  $71\pm80 (0.3)$  &  $1.13\pm0.06 \,(2.8)$ & $0.10\pm0.08 \,(0.3)$ \\
     5 & 2.2 & 11.2 & $2.39\pm0.63 \,(0.5) $ & $6.8\pm7.0 \,(0.4) $ & $---$     & $---$             &  $1.30\pm0.14 \,(0.9) $ & $0.33 \pm 0.31 \,(0.5)$ \\
    \hline \hline
  \end{tabular}
\end{table*}

\section{Application to the 2\lowercase{d}FGRS}

In this Section we describe the construction of volume limited samples
from the 2dFGRS (Section 3.1) and outline how we deal with the small,
remaining incompleteness of the survey when we measure the CPDF
(Section 3.2). The estimation of errors on the measured higher order
moments is described in Section 3.3.
We use the full 2dFGRS as our starting point. The final spectra were
taken in April 2002, giving a total of 221,414 galaxies with high
quality redshifts (i.e. with quality flag Q $\ge3$; see Colless et
al. 2001).  The median depth of the full survey, to a nominal
magnitude limit of $b_{\rm J} \sim 19.45$, is $z \sim 0.11$. We
consider the two large contiguous survey regions, one near the south
galactic pole (SGP) and the other towards the north galactic pole
(NGP).  After restricting attention to the high redshift completeness
parts of the survey (see Colless et al. 2001; Norberg et al. 2002b),
the effective solid angle covered by the NGP region is 469 square
degrees and that of the SGP is 670 square degrees.  Full details of
the 2dFGRS and the construction and use of the mask quantifying the
completeness of the survey can be found in Colless et al. (2001,
2003).

We make use of mock 2dFGRS catalogues to test our algorithm for dealing 
with the spectroscopic incompleteness of the survey and to estimate 
errors on the measured higher order correlation functions. 
The construction of the mocks is described in Norberg et al. (2002b). 
In short, catalogues are extracted from the Virgo Consortium's 
$\Lambda$CDM Hubble Volume simulation which covers a volume of 
$27{\rm Gpc}^{3}$ (Evrard et al. 2002). 
A heuristic bias scheme is applied to the smoothed distribution 
of dark matter in the simulation to select `galaxies' with a 
specified clustering pattern (Cole et al. 1998). The parameters 
of the biasing scheme are adjusted so that the extracted galaxies 
have the same projected correlation function as measured for the 
flux limited 2dFGRS by Hawkins et al. (2003).
Observers are placed within the Hubble Volume simulation according to 
the criteria set out in Norberg et al. (2002b). 
Mock catalogues are then extracted by applying the radial and 
angular selection functions of the 2dFGRS. Finally, the mock is 
degraded from uniform coverage within the angular mask of the survey by 
applying the spectroscopic completeness mask of the 2dFGRS.

\subsection{Construction of volume limited catalogues}

In a flux limited sample the density of galaxies is a strong function
of radial distance. This effect needs to be taken into account in
clustering analyses (for an example of a technique appropriate to a
counts-in-cells analysis, see Efstathiou et al. 1990).  Alternatively,
one may construct volume limited samples in which the radial selection
function is constant and any variations in the density of galaxies are
due only to large scale structure.  This greatly simplifies the
analysis at the expense of using a subset of the survey galaxies. The
2dFGRS contains enough galaxies and covers sufficient volume to permit
the construction of volume limited samples corresponding to a wide
baseline in luminosity from which robust measurements of the higher
order correlation functions can be obtained.

We follow the approach taken by Norberg et al. (2001, 2002a) who
measured the projected 2-point correlation function of 2dFGRS galaxies
in volume limited samples corresponding to bins in absolute magnitude.  
The samples are defined by a specified absolute magnitude range, with
absolute magnitudes corrected to zero redshift (this correction is
made using the $k+e$ correction given by Norberg et al. 2002b).  As
any survey has, in practice, a bright as well as a faint flux limit,
this implies that a selected galaxy should fall between a minimum
($z_{\rm min}$) and a maximum ($z_{\rm max}$) redshift.  This then
guarantees that all sample galaxies are visible within the flux limits
of the survey when displaced to any depth within the volume of the
sample.  The properties of the combined NGP and SGP volume limited
samples examined in this paper are given in Table~1.

\subsection{Correcting for incompleteness}

\begin{figure}
\plotone{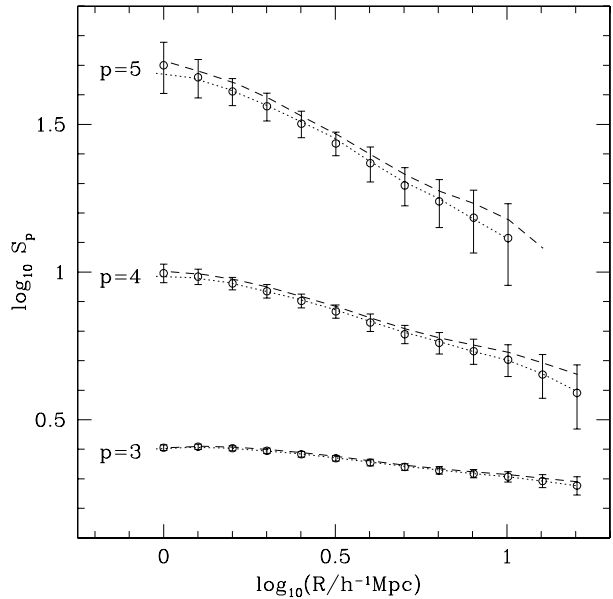}
\caption{ 
A test of the scheme used to correct the measured distribution of 
counts-in-cells for incompleteness in the 2dFGRS, using mock catalogues. 
The plot shows the hierarchical amplitudes, $S_p$, for orders $p=3,4$ and 
$5$. The dotted lines show the results from fully sampled mocks that 
have no incompleteness. The dashed lines show how these results change 
when the completeness mask of the 2dFGRS is applied to the mocks and 
{\it no} compensation is made for the variable spectroscopic completeness. 
The circles show the $S_p$ recovered once the correction to the cell 
radius discussed in the text is made. The errorbars show the {\it rms} 
scatter estimated from the mock catalogues.
}
\label{fig:test}
\end{figure}

There are two possible sources of incompleteness that need to be
considered when estimating the galaxy count within a cell.  The first
is volume incompleteness, which can arise when some fraction of the
cell volume samples a region of space that is not part of the 2dFGRS.
This situation can arise because the survey has a complicated boundary
and also because it contains holes excised around bright stars and
other interlopers in the parent APM galaxy catalogue (Maddox et
al. 1990, 1996).  
The second source of incompleteness is spectroscopic incompleteness.
The final 2dFGRS catalogue is much more homogeneous than the 100k
release (contrast the completeness mask of the final survey shown in
figure~1 of Hawkins et al. 2003 with the equivalent mask depicted in
figure~15 of Colless et al. 2001.)  However, the spectroscopic
completeness still varies with position on the sky and needs to be
incorporated into the counts-in-cells analysis.

It is therefore necessary to devise a strategy to compensate for the
fact that a cell will sample regions that have varying spectroscopic
completeness and which may even straddle the survey boundary or a
hole.  We project the volume enclosed by the cell onto the sky and
estimate, using the survey masks, the mean combined spectroscopic and
volume completeness, $f$, within the sphere.  Rather than view the
consequence of this incompleteness as missed galaxies, we instead
consider it as missed \emph{volume}. We compute a new radius
for the sphere given by $R' = f^{-\frac{1}{3}} R$: such a sphere with
radius $R'$ will have an incomplete volume equivalent to that of a
fully complete sphere of radius $R$.  Spheres for which $f$ is less
than $50\%$ are discarded.  The galaxy count within the sphere of
radius $R'$ then contributes to the CPDF at the effective radius $R$.
Each sphere thrown down is individually scaled in this way according
to its local incompleteness, as given by the survey masks.  We note
that, due to our chosen acceptable minimum completeness of $50\%$, the
rescaling of the cell radius is always less than the width of the
radial bins we use to plot the higher order correlation functions.
Our results are insensitive to the precise choice of completeness
threshold.

An alternative method to correct cell counts is described in
Efstathiou et al. (1990).  In this commonly used approach it is the
galaxy \emph{counts} which are scaled up in proportion to the degree
of incompleteness in the cell, as opposed to the cell volume as we
have done.  We have tried both correction methods when calculating the
higher order moments and find the results are essentially identical
(see Croton et al. 2004 for further discussion of the relative
strengths and weaknesses of both methods).

A test of our method for dealing with incompleteness is shown in
Fig.~\ref{fig:test}. This plot shows the $S_p$ estimated from the
higher order correlation functions measured in mock 2dFGRS catalogues
(Norberg et al. 2002b).  The dotted lines show the results for complete
mocks, with uniform sampling of the galaxy distribution within the
full angular boundary of the 2dFGRS.  The dashed lines show how these
results change once the mocks are degraded to mimic the spectroscopic
incompleteness and irregular geometry of the 2dFGRS, without applying
any correction to compensate for this incompleteness.  The circles
show the values of $S_p$ recovered on application of the correction
for incompleteness described above. These results are in excellent
agreement with those from the fully sampled, `perfect' mocks.

We have carried out two independent counts-in-cells analyses, using
different algorithms to place cells within the survey volume. The
results are insensitive to the details of the counts-in-cells
algorithm.  The CPDF is measured using $2.5 \times
10^7$ cells for each cell radius.  We have further checked the counts
in cells analysis by comparing the measured two point volume averaged
correlation function with the integral of the measured spatial two
point correlation function, given by Eq.~\ref{volxi}; the integral of
the spatial correlation function is in very good agreement with the
direct estimate of the volume averaged correlation function.

\subsection{Error Estimation}
\label{s:errors}

\begin{figure}
\plotone{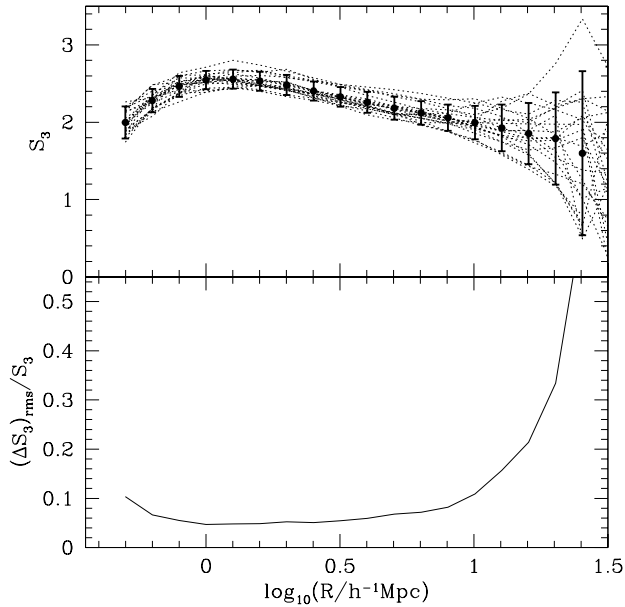}
\caption{ The upper panel shows the skewness, $S_3$, recovered 
from mock 2dFGRS catalogues in volume limited samples defined 
by the magnitude range $-19>M_{b_{\rm J}}-5\log_{10}h>-20$. 
The dotted lines show the skewness measured in each catalogue. 
The points show the mean skewness. The errorbars show the mean {\it rms} 
scatter averaged over 22 mocks, as described in the text in Section 
\ref{s:errors}.  The lower panel shows the fractional 
error as a function of cell radius. 
This panel shows how well we can expect to measure the skewness in 
a catalogue of this size extracted from the 2dFGRS, including the 
contribution from sampling variance.}
\label{fig:mocks}
\end{figure}

We estimate the error on the higher order correlation functions and
hierarchical amplitudes using the set of 22 mock 2dFGRS surveys described 
by Norberg et al. (2002b).  
The $1 \sigma$ errors that we show on plots 
correspond to the \emph{rms} scatter 
over the ensemble of mocks (see Norberg et al. 2001a).  
To recap, we consider one of the mocks as the
``data'' and compute the variance around this ``mean'' using the
remaining mock catalogues.  This process is repeated for each mock in
turn, and the {\it rms} scatter is taken as the mean variance.  We illustrate
this approach in Fig.~\ref{fig:mocks} for the case of $p=3$, for a
volume defined by the magnitude range $-19>M_{b_{\rm
J}}-5\log_{10}h>-20$.  In the upper panel, the skewness or $S_3$
measured in each mock is shown by the dotted lines. The points show
the mean skewness averaged over the ensemble of 22 mocks. The
errorbars show the {\it rms} scatter on these measurements.  The lower
panel shows the fractional error that we expect on the measurement of
$S_3$ for this particular volume limited sample. Beyond $20h^{-1}$Mpc,
the fractional error increases rapidly. Our estimate of the fractional
error automatically includes the contribution from sampling
variance due to large scale structure (sometimes referred to as
``cosmic variance'').
To estimate the error on a measured correlation function, we
simply compute the fractional {\it rms} scatter for the equivalent
volume limited sample using the ensemble of mocks, and multiply the
measured quantity by the fractional error.

We have compared the estimate of the {\it rms} scatter from the
ensemble of mocks with an internal estimate using a jackknife
technique (see, for example, Zehavi et al. 2002). In the jackknife
approach, the survey is split into subsamples. The error is then the
scatter between the measurements when each subsample is omitted in
turn from the analysis.  The jackknife gives comparable errors to the
mock ensemble for low order moments. On large scales, the higher order
moments are particularly sensitive to sample variance and, in these
cases, the jackknife approach can only provide a lower bound to the
true scatter.

A more formal error estimation procedure is adopted 
when computing the best fit values for the hierarchical 
amplitudes, $S_p$. In this case, we employ a principal component 
analysis to explicitly take into account the correlations between 
the $S_p$ inferred at different cell radii (see e.g. Porciani \& 
Giavalisco 2002 and Section 6 of Bernardeau et al. 2002). 
The mock catalogues are used to compute the full 
covariance matrix of the $S_p$ data points to be fitted. 
Next, the eigenvalues and eigenvectors of the covariance matrix 
are determined. We find that, typically, the first few eigenvectors are 
responsible for over $90\%$ of the variance. Given the number of data 
points that we consider in the fits, this means that we have 
around a factor of two to three times fewer independent 
points than data points fitted. (Details of the range of scales 
used in the fits will be given in Section 4.2.)
We note that in most previous work, the $S_p$ measured at different cell 
radii were simply averaged together ignoring {\it any} correlations 
between bins, resulting in unrealistically small errors in the fitted values.

\begin{figure}
\plotone{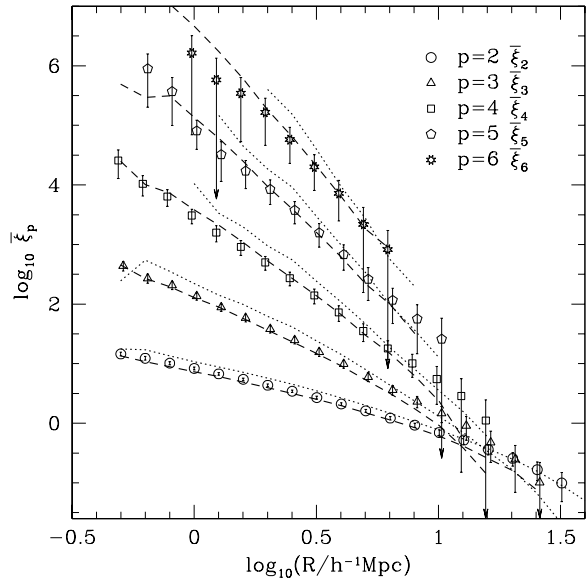}
\caption{ 
The higher order correlation functions measured for 2dFGRS galaxies. 
The symbols show the measurements for galaxies in the absolute magnitude 
range $-19>M_{b_{\rm J}}-5\log_{10}h>-20$; the key gives the order $p$. 
The lines show the results for different luminosity samples; the dashed 
lines show the $\bar{\xi}_{p}$ for galaxies with 
$-18>M_{b_{\rm J}}-5\log_{10}h>-19$ and the dotted lines 
show the results for galaxies with $-20>M_{b_{\rm J}}-5\log_{10}h>-21$.
}
\label{fig:xip}
\end{figure}

\begin{figure}
\plotone{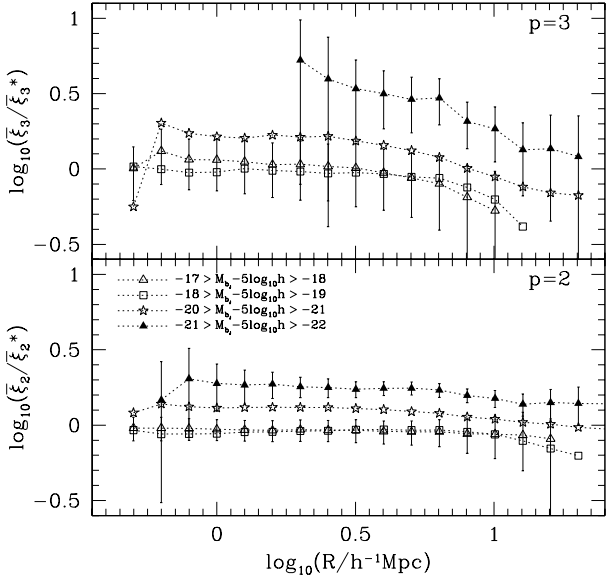}
\caption{ 
The dependence of the higher order correlation functions on luminosity. 
The orders $p=3$ (top panel) and $p=2$ (bottom panel) are shown. 
The correlation functions for samples of different luminosity are divided 
by the correlation function measured for $L_*$ galaxies, with 
$-19>M_{b_{\rm J}}-5\log_{10}h>-20$.
}
\label{fig:xipratio}
\end{figure}

\section{Results}

\subsection{Volume-averaged correlation functions}

The volume averaged correlation functions estimated from the CPDF
constructed from the combined NGP and SGP cell counts are plotted in
Fig.~\ref{fig:xip}.  The symbols show the correlation functions for 
the $L_*$ sample with $-19>M_{b_{\rm J}}-5\log_{10}h>-20$. The lines show 
the measurements made for galaxies in magnitude bins adjacent to $L_*$ (the 
dashed lines correspond to a sample that is one magnitude fainter and the 
dotted lines to a sample that is one magnitude brighter).
The correlation functions steepen dramatically on small scales as the 
order $p$ increases.  

To better quantify the dependence of the higher order correlation functions on 
luminosity, we plot the ratio of the $\bar{\xi}_{p}$ to the results for 
the $L_*$ reference sample in Fig.~\ref{fig:xipratio}, for  
the cases $p=2$ and $p=3$.  
The variance in the distribution of counts-in-cells on a given
smoothing scale increases with the luminosity of the volume limited
sample (see the bottom panel of Fig.~\ref{fig:xipratio}).  This effect is
similar to that reported by Norberg et al. (2001, 2002a), who measured
the dependence of the strength of galaxy clustering on luminosity in real
space, whereas our results are in redshift space. This behaviour is
broadly seen to extend to the higher order clustering, however
the ranking of the amplitude of the higher order correlation functions
with luminosity is not always preserved on large scales. 
This issue is investigated further in Section \ref{s:system}.

\subsection{Hierarchical clustering}

\begin{figure}
\plotone{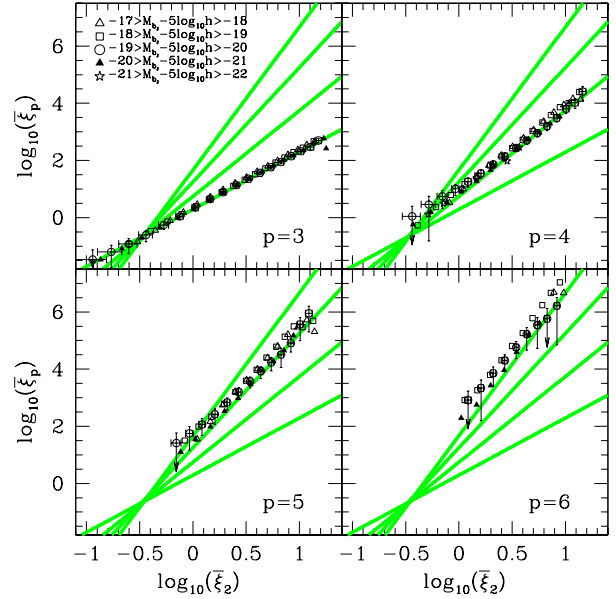}
\caption{
The volume averaged correlation functions, 
$\xibar_p$, for $p=3$ to $6$, plotted as a function of the
variance, $\xibar_2$. Each panel corresponds to a different order 
plotted on the ordinate, as indicated by the legend. 
(Note that $\bar{\xi}_{5}$ and $\bar{\xi}_{6}$ are not plotted 
for the brightest sample, as they are too noisy.) 
The symbols refer to different magnitude ranges as given by the 
key in the first panel. 
The line styles denote the results for different absolute
magnitude ranges, as indicated by legend.  
The thick grey lines show power-laws with slopes of 
$2$, $3$, $4$ and $5$ in order of increasing amplitude, which 
are intended to act as a reference.  
}
\label{fig:hier}
\end{figure}

We use the measured volume averaged correlation functions from
Fig.~\ref{fig:xip} to test the hierarchical clustering model set out
in Section 2.2 and Eq.~\ref{eq:sp}.  In Fig.~\ref{fig:hier}, we plot
the $p=3$--$6$ point volume averaged correlation functions as a
function of the variance (or two-point function) measured on the same
scale.  Small values of the moments correspond to large cells. The
thick grey lines show the higher order moments expected in the
hierarchical model.  (From Eq.~\ref{eq:sp}, the offsets of these lines
are the hierarchical amplitudes $S_p$. We have used the best fit
values of $S_p$ that we obtain later on in this Section. However, the
width of the lines does not indicate the error on the fit: the lines
are intended merely to guide the eye.)  On small scales (large
variances), hierarchical scaling is followed.  On intermediate and
large scales, for which the variance drops below $\sim 1.3$, the
measured moments depart somewhat from the hierarchical scaling
behaviour, particularly in the case of the higher orders.

\begin{figure*}
\plotfull{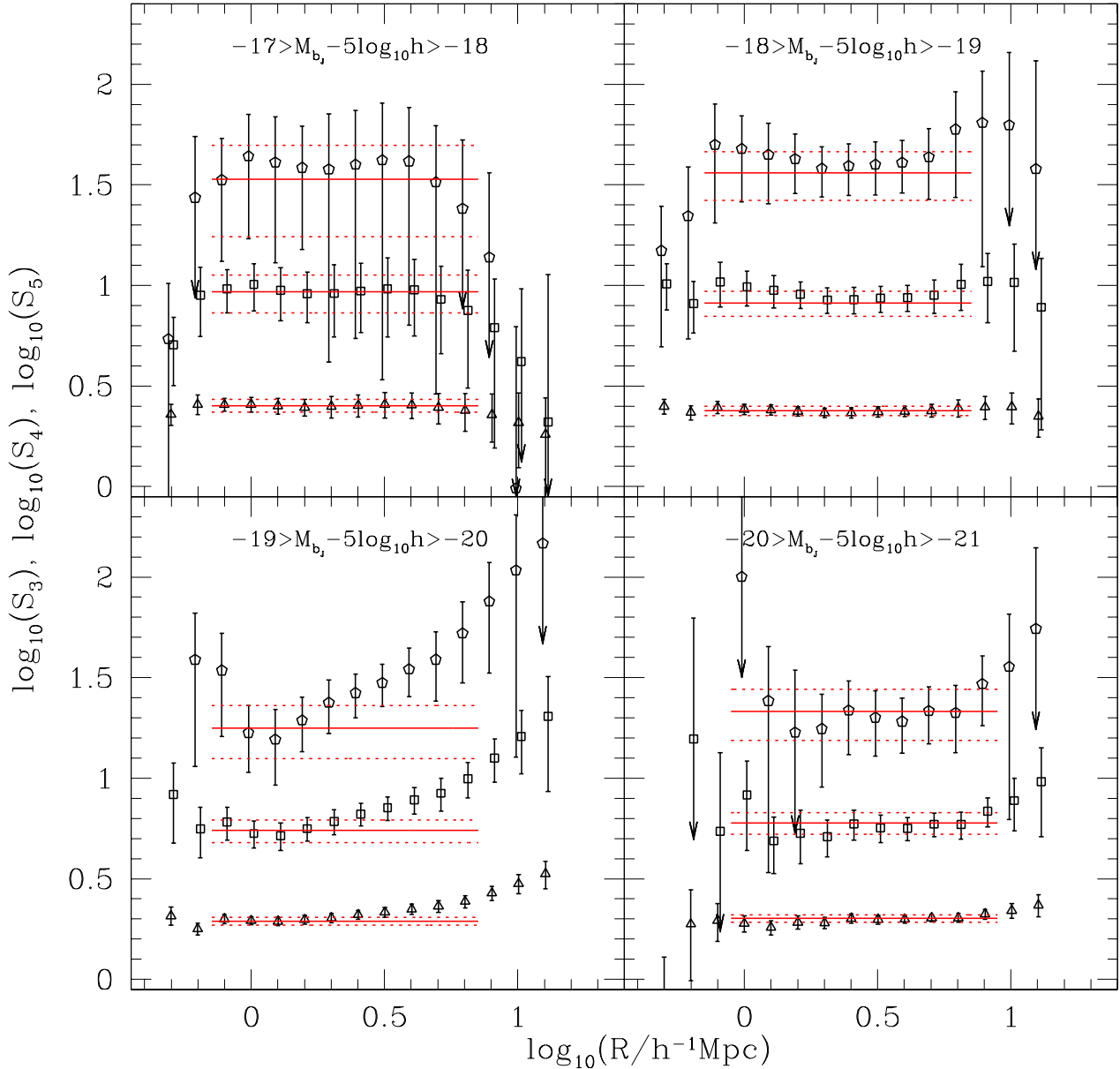}
\caption{
The hierarchical amplitudes, $S_p$, for $p=3,4$ and $5$, plotted 
as a function of cell radius for the galaxy samples defined in 
Table~1. Each panel shows the results for a different volume 
limited catalogue, as indicated by the legend. 
The points with errorbars show the results obtained from the 
full volume limited samples: triangles show $S_3$, squares show 
$S_4$ and pentagons show $S_5$.
The solid lines show the best fit values and the dotted lines 
indicate the $1\sigma$ errors on the fits, as described in the 
text. The lines are plotted over the range of scales used in the 
fits. 
}
\label{fig:sp}
\end{figure*}

The hierarchical scaling of the higher order correlation functions is
exploited to plot the hierarchical amplitudes $S_{p} = \xibar_p /
\xibar_2^{p-1}$ as a function of cell radius in
Fig.~\ref{fig:sp}. Each panel corresponds to a different volume
limited sample, where the lines and points correspond to
$S_3$, $S_4$ and $S_5$ in order of increasing amplitude.  The
hierarchical amplitudes measured from the two brightest volume limited
samples systematically show an increase around $10h^{-1}$Mpc.  This
effect is particularly significant in the $-19 > M_{b_{\rm J}}-5 \log_{10}
h > -20$ sample, with the $S_p$ increasing by a factor of 2 to 5
depending on $p$.  
On smaller scales the
hierarchical amplitudes are essentially independent of the cell radius
for all magnitude ranges considered.  It should be noted that the
$S_p$ measured in real space vary more strongly in amplitude with
scale than in redshift space, particularly at small cell radii 
(Gazta\~{n}aga 1994; Szapudi et al. 1995, Szapudi \& Gazta\~naga 1998).

We have fit constant values to the measured $S_p$, using the principal
component analysis outlined in Section 3.3. This approach takes into
account the correlations between the measurements on different
scales. The range of scales used to fit $S_p$ is held fixed for each
volume limited sample and is quoted in Table~2. Typically, there are
ten values of $S_p$ in the range considered in the fits. The principal
component analysis reveals that just $2-4$ linear combinations of
these points account for more than $90$\% of the variance; this gives
a fairer impression of the number of independent data points.  The
principal eigenvector is in all cases almost independent of scale, i.e. its
effect is to move all the points coherently up and down (driven by
large scale variation in the mean density estimated from the survey). 
Therefore, the
best fitting constant tends to favour a fit either slightly above or
below each set of data points. This is exactly what is seen in the
various panels of Fig.~6.  The best fit constants to the measured
$S_p$ are given in Table~2, along with an error from the principal
component analysis.  The fits to $S_3$ and $S_4$ for the $-19 >
M_{b_{\rm J}}-5 \log_{10} h > -20$ sample are poor in terms
of the reduced $\chi^2$.  There some dependence
of the $S_p$ with increasing luminosity. This behaviour is explored in Section~5.

\subsection{Systematic effects: the influence of superclusters}

\label{s:system}

The higher order moments of the CPDF are sensitive to the presence of
massive structures that contribute to the extreme event tail of the
count distribution. It is therefore important to examine the 2dFGRS to
look for any rare large scale structures that could exert a significant 
influence on the form of the CPDF.  The projected density of galaxies in
the right ascension-redshift plane for a volume limited catalogue 
defined by the magnitude range $-19 > M_{b_{\rm J}}-5 \log_{10} h > -20$ 
is plotted in figure~1 of Baugh et al. (2004). 
There are two clear hot spots or superstructures apparent in this figure, 
one in the NGP at a redshift of $z=0.08$ and a right ascension of 3.4 
hours, and the other in the SGP at $z=0.11$ at a right ascension 
of $0.2$ hours.  These structures are confirmed as superclusters 
of galaxies in the group catalogue constructed from the flux limited 
2dFGRS (Eke et al. 2004); of the 94 groups in the full flux limited 
survey out to $z\sim 0.15$ with 9 or more
members and estimated masses above $5 \times 10^{14} h^{-1}M_{\odot}$,
20\% reside in these superclusters.  As a result of the redshift at
which these superclusters lie, these structures are only influential
in volume limited samples brighter than $M_{b_{\rm J}}-5 \log_{10} h =
-18$.

The results presented earlier in this Section show features that could 
be due to the presence of these superclusters. 
For example, the volume averaged correlation functions for 
the $-19 > M_{b_{\rm J}}-5 \log_{10} h > -20$ sample plotted in 
Fig.~\ref{fig:xip} appear to have more power on large scales than those 
measured from the other volume limited samples. 
This is consistent with the theoretical expectations
for measurements that are strongly affected by the
presence of a supercluster: a boost in the clustering amplitude on
large scales, due to a structure with a larger bias, and a reduction
in the clustering amplitude on small scales arising from the large
velocity dispersions within the clusters making up the structure.

To investigate this hypothesis, we have carried out the test of 
removing the two superclusters from the sample and recomputing 
the volume averaged correlation functions. The goal of this exercise 
is not to ``correct'' the measured correlation functions but rather 
to illustrate the impact of the superclusters on our results. 
We remove the superclusters by masking out their central densest 
regions, corresponding to prohibiting the placement of cells 
within a sphere of radius $25h^{-1}$Mpc from each supercluster 
centre (for a different approach on how to take this 
type of effect into account see Colombi, Bouchet \& Schaeffer 1994 
and Fry \& Gazta\~naga 1994).

\begin{figure}
\centering{\plotone{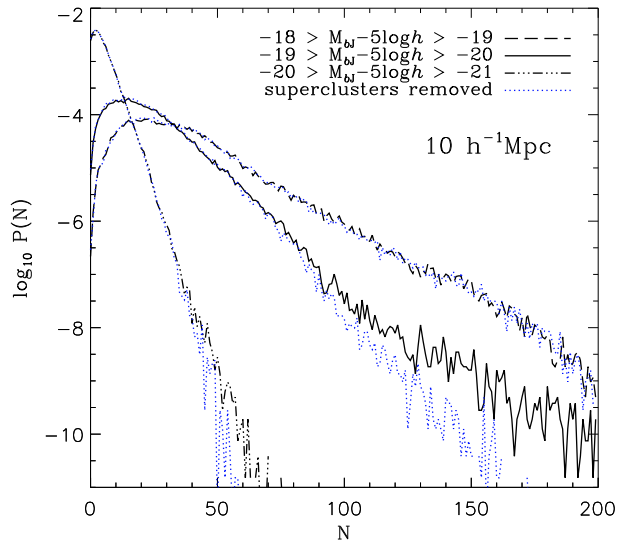}}
\caption{ 
The probability, $P_N$, of finding exactly $N$ galaxies in randomly placed 
cells of radius $10h^{-1}$Mpc (the CPDF, Eq.~2), for different volume limited 
galaxy samples.  Each bold line shows the full volume CPDF, while the 
individual dotted lines give the result after the supercluster regions have 
been omitted from the analysis, as described in Section~4.3.}
\label{fig:PN} 
\end{figure}

Fig.~\ref{fig:PN}  shows the effect of the supercluster removal on the tail of the CPDF 
for $10h^{-1}$Mpc radius cells, calculated for three volume limited 
catalogues centred on $L^*$.  The mean number of galaxies in a cell for each 
galaxy sample is roughly $40$, $24$, and $6$ going from faintest to brightest.  
The presence of the two superclusters makes a clear difference to the high $N$ 
counts for galaxy samples brighter than $M_{b_{\rm J}}-5\log_{10}h =-19$.  The 
maximum redshift of the faint volume limited catalogue in this
figure only  marginally includes the NGP supercluster, and
so $P_N$ remains essentially unaffected in this case.

\begin{figure}
\plotone{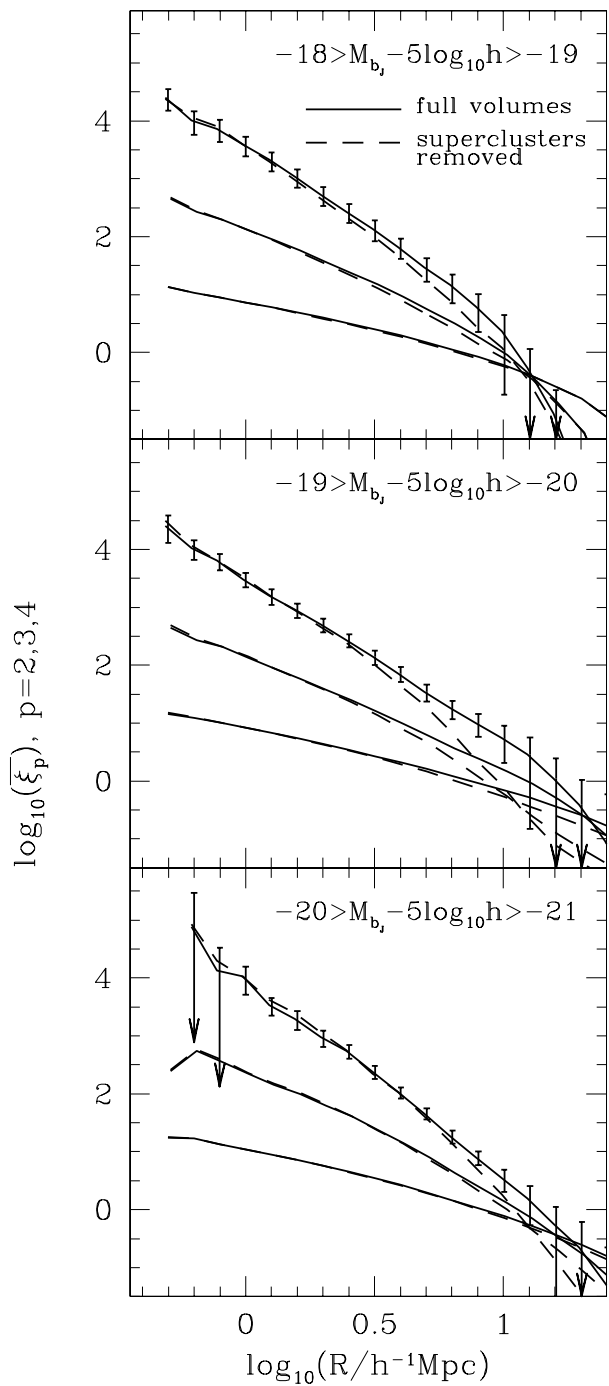}
\caption{ 
The volume averaged correlation functions for $p=2$ to $4$, with each 
panel showing the results from a different volume limited sample, as indicated 
by the legend.  The solid lines show the estimates from the full volumes 
and the dashed lines show the results when the supercluster regions 
are omitted from the analysis.  For clarity, errorbars are only plotted 
on the solid curves for order $p=4$.}
\label{fig:xip2} 
\end{figure}

Fig.~\ref{fig:xip2} shows volume averaged correlation functions of order
$p=2,3,4$ for three volume limited catalogues from Table~1, where each
panel corresponds to a fixed absolute magnitude range.  The lines
correspond to different orders of clustering, starting with the lowest
in amplitude, the two point volume averaged correlation function, and
moving through to the four point function, at which we stop plotting the 
results  for clarity although the trends shown continue up to sixth
order. The solid curves show the correlation functions measured from
the full volume limited samples, as shown previously in Fig.~6, and the dashed lines show the results
when the regions containing the superclusters are excluded from the
CPDF.  The higher order correlation functions are systematically
boosted on intermediate and large scales when the superclusters are
included in the analysis. The precise scale on which the correlation
functions become sensitive to the presence of the superclusters
depends upon the order; for the four point function, the two estimates
of the correlation function typically deviate for cells of radius
$3h^{-1}$Mpc and larger.

\begin{figure}
\plotone{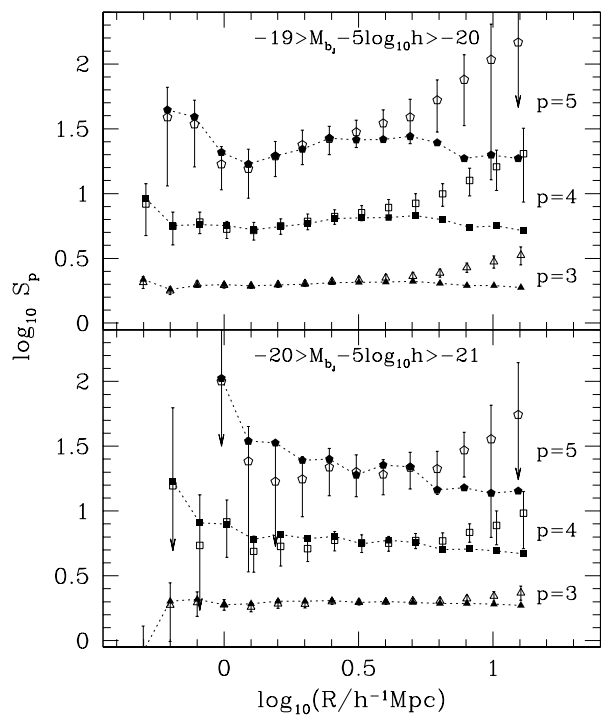}
\caption{
The hierarchical amplitudes, $S_3$ (triangles), $S_4$ (squares)
and $S_5$ (pentagons).
The top panel corresponds to galaxies with 
$-19>M_{b_{\rm J}}-5\log_{10}h>-20$ and the bottom panel 
to $-20>M_{b_{\rm J}}-5\log_{10}h>-21$.
The open symbols with errorbars show the results obtained using the 
full volume limited catalogues. The filled symbols show
 how the results change when regions containing 
the two superclusters are omitted from the analysis.
}
\label{fig:sp2}
\end{figure}

The impact on the hierarchical amplitudes, $S_{p}$, of removing the 
superclusters in shown in Fig.~\ref{fig:sp2}, in which we plot the 
results for the volume limited sample defined by 
$-19>M_{b_{\rm J}}-5\log_{10}h>-20$.
In Fig.~\ref{fig:sp2}, the open points show the 
hierarchical amplitudes measured from the full volume limited 
sample. The filled symbols show the results 
obtained from the same volume but with the supercluster regions 
masked out. 
The $S_p$ obtained when the two superclusters are removed from the
analysis are much closer to being independent of cell size.  The
sensitivity of higher orders to rare peaks
has been noticed in earlier analyses of galaxy surveys
(Groth \& Peebles 1977; Gazta\~{n}aga 1992; Bouchet et al. 1993; 
Lahav et al. 1993; ; Gazta\~{n}aga 1994; Hoyle et al. 2000).

\section{Interpretation and the implications for galaxy bias}

In this Section we quantify how the hierarchical amplitudes scale with
galaxy luminosity and discuss the implications of our results for
simple models of galaxy bias.  We first test the hypothesis set out in
Section 2.4 that the variation in clustering with luminosity apparent
in Fig.~\ref{fig:xip} can be described by a single, relative bias
factor, as defined by Eq.~\ref{eq:br}.  The relative bias factors, $b_r$,
computed from the variance and the deviation from the linear bias
model, as quantified by $c_2'$ (Eq.~\ref{eq:c2'}), are listed in
Table~2; here the mean value is given by the best $\chi^2$ fit over
all cell radii.  The change in the amplitude of the relative bias with
sample luminosity, shown in Fig.~4, is in excellent agreement with the trend found by
Norberg et al. (2001), who analysed the projected spatial clustering
of 2dFGRS galaxies.  This agreement is remarkable given the different
approaches used to measure the two-point correlations and the fact
that the analysis in this paper is in redshift space, whereas the
study carried out by Norberg et al. was unaffected by peculiar
motions.

The coefficients $c_2'$ are different from zero at a 1-$\sigma$ level.
These findings are consistent with a small deviation from the
linear biasing model (at a 2-sigma level for the brighter
samples). This is in qualitative agreement with the estimation
of $c_2$ using the the bispectrum (Scoccimarro 2000, Verde et al. 2002) or
the 3-point function measured from the parent APM galaxy survey
(Frieman \& Gazta\~naga 1999).

\begin{figure}
\plotone{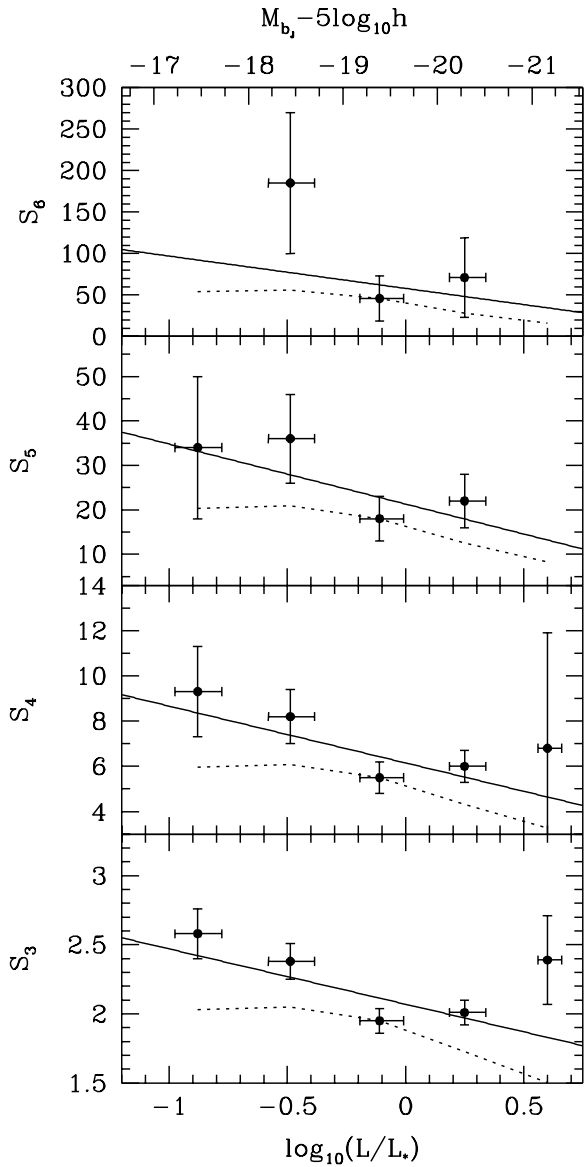}
\caption{ 
The variation of the hierarchical amplitudes, $S_p$, with 
absolute magnitude. 
The points are plotted at the median magnitude of each volume 
limited sample and the horizontal bars indicate the interval in 
which $25\%$ to $75\%$ of the galaxies fall, computed using 
the 2dFGRS luminosity function fit quoted by Norberg et al. (2002b).
Each panel shows the results for a different order of clustering. 
The dotted line shows predictions of the linear relative bias model 
for the variation of the $S_p$ with luminosity (Eq.~15). The solid lines show 
linear fits in log luminosity to the observed trend in the value of $S_p$ 
with sample luminosity (see text of Section $5$ for details).
}
\label{fig:spl}
\end{figure}

The variation of the hierarchical amplitudes with luminosity is
plotted in Fig.~\ref{fig:spl}.  Each panel corresponds to a different
order $p$. The filled points show the hierarchical amplitudes
averaged over the different cell radii employed (these values and the
associated errors are given in Table~2).  The dotted line shows the
hierarchical amplitudes predicted by the linear relative bias model 
(Eq.~15), using the best fit bias factors stated in Table~2.  This model
gives a rough approximation to the data.  However, the observed
variation of $S_p$ with luminosity is somewhat better described by a
linear fit in the logarithm of luminosity, as shown by the solid
lines.  This implies that the dependence of the hierarchical
amplitudes on luminosity is more complicated than expected in the
simple relative bias model of Eq.~\ref{eq:SJlinear} (as does the 
fact that we find some evidence for non-zero values for $c_2'$).  
The solid lines show the best linear fit to the hierarchical 
amplitudes as a function of the logarithm of the median luminosity 
of the samples:
\begin{equation} 
S^{G}_{p} = A_{p} + B_{p} \log_{10}\left(\frac{L}{L_{*}}\right).
\end{equation} 
We find a greater than $2-\sigma$ ($\Delta \chi^2 > 7.2$ for two parameters) 
detection of a non-zero value for $B_3$. 
However, for $p>3$, the constraints on $B_p$ are much weaker and 
there is no clear evidence for a luminosity dependence in the $S_p$ 
values in these cases. 
For completeness, the best fit values for each order are: 
$(A_{3}, B_{3}) = (2.07, -0.40)$,  
$(A_{4}, B_{4}) = (6.15, -2.51)$,  
$(A_{5}, B_{5}) = (21.3, -13.5)$,  
$(A_{6}, B_{6}) = (58, -39)$.


\section{Conclusions}

In this paper we have measured the higher order correlation functions
of galaxies in volume limited samples drawn from the 2dFGRS.  The most
recent comparable work is the analysis of the Stromlo-APM and UKST
redshift surveys by Hoyle, Szapudi \& Baugh (2000). These authors also
considered volume limited subsamples drawn from the flux limited
redshift survey.  The largest UKST sample considered by Hoyle et
al. contained 500 galaxies and covered a volume of $9 \times 10^{5}
h^{-3}{\rm Mpc}^{3}$; the reference sample used in our work contains
90 times this number of galaxies and covers ten times the volume. In
our analysis, we can follow the variation of clustering over more than
a decade in luminosity, whereas Hoyle et al. had to focus their
attention around $L_*$.

The measurement of the higher order galaxy correlation functions is
still challenging, however. In spite of the order of magnitude
increase in size that the 2dFGRS represents over previously completed
surveys, we have found that the higher order moments that we measure
are somewhat sensitive to the presence of large structures. In
particular, there are two superclusters that influence our
measurements, one in the SGP region and the other in the NGP. These
structures contain a sizeable fraction of the cluster mass groups in
the 2dFGRS (Eke et al. 2004).  The inclusion of these structures has
an impact on our estimates of the three point and higher order volume
averaged correlation functions on scales around $4-10h^{-1}$Mpc and
above, depending on the order of the correlation function. For this
reason, we have presented measurements of the higher order correlation
functions both with and without these structures. We stress that
the removal of these superclusters should not be considered a
correction to the full catalogue results, but rather as an indication 
of the impact of rare structures on our results for the higher order 
moments.
On the other hand, the up-turn that we find in the values of the 
hierarchical amplitudes on large scales is predicted by some 
structure formation models; for example models with non-Gaussian 
initial density fields predict a similar form for the $S_p$ as we 
measure from the full volume limited samples 
(Gazta\~{n}aga \& M\"{a}h\"{o}nen 1996; Gazta\~{n}aga \& Fosalba 1998; 
Bernardeau et al. 2002).

The difficulties in estimating $S_p$ values on large, quasi-linear
scales ($> 10 h^{-1}$Mpc), prevent a direct comparison with
perturbation theory (see Bernardeau et al. 2002). The current best estimates
on these scales
are still those measured from the angular APM Galaxy Survey
(Gazta\~naga 1994, Szapudi et al. 1995, Szapudi \& Gazta\~naga 1998). 
At the time of writing, the results from the SDSS Early Data Release 
are still limited to small scales (Gazta\~naga 2002, Szapudi et al. 2002).
Despite being unable to make a robust measurement of the higher order
correlation functions on the very large scales for which weakly non-linear
perturbation theory is applicable, we are still able to reach a number
of interesting conclusions:

\begin{itemize} 

\item[(i)] We have demonstrated that the higher order galaxy
correlation functions measured from the 2dFGRS follow a hierarchical
scaling.  Baugh et al. (2004) showed that $L_*$ galaxies display 
higher order correlation functions that scale in a hierarchical 
fashion; we have extended these authors' analysis to cover a wide 
range of galaxy luminosity.
The higher order moments of the galaxy count distribution
are proportional to the variance raised to a power that depends upon
the order of the correlation function under consideration.
This behaviour holds on physical scales ranging from those on which we expect
the underlying density fluctuations to be strongly nonlinear all the
way through to quasi-linear scales.  This scaling has been tested up
to the six point correlation function for the first time using a
redshift survey.  This confirms the conclusions of a complementary
analysis carried out by Croton et al. (2004), who found
hierarchical scaling when measuring the reduced void probability
function of the 2dFGRS.

\item[(ii)] We have estimated values of the hierarchical amplitudes,
$S_p = \xibar_{p}/\xibar^{p-1}_{2}$, for cells of different radii. The
hierarchical amplitudes are approximately constant on small to medium
scales (depending on the order considered), while for the larger
volumes, $S_p$ seem to increase with radius at large scales. Although
this could in principle result from a boundary or mask effect
(e.g. see Szapudi \& Gazta\~naga 1998; Bernardeau et al. 2002), we
have shown with mock catalogues that this is not the case here
(e.g. see Fig.~1). If the two most massive superclusters in the survey
are removed from the analysis, the hierarchical amplitudes are
remarkably independent over all scales.  That  the $S_p$ are roughly
constant on small scales, with smaller amplitudes than in real space
(e.g. Gazta\~naga 1994), has been noted before for measurements in
redshift space. It arises due to a cancellation of the enhanced signal
on small scales in real space by a damping of clustering in redshift
space due to peculiar motions (Lahav et al. 1993; Fry \& Gazta\~naga
1994; Hivon et al. 1995; Hoyle et al. 2000; Bernardeau et al. 2002).

\item[(iii)] We find that the amplitude of the higher order correlation
functions scales with luminosity. The magnitude of the luminosity segregation 
increases with the order of the correlation (see Fig.~4).
For the variance, $\xibar_2$,
the strength of the trend is in very good agreement with that reported by
Norberg et al. (2001), but note that these authors measured the luminosity
segregation in real space, whereas our results are in redshift space.
The strength of the luminosity segregation for higher orders can be 
mostly explained as the result of hierarchical scaling 
$\xibar_{p} \sim \xibar_2^{p-1}$, so that most of the effect can be 
attributed to luminosity segregation in the variance. This can be seen in
Fig.~5 where data from different luminosities trace out the
same hierarchical curve with little scatter.

\item[(iv)]  
We find some evidence for a residual dependence of $S_p$ on 
luminosity, although the effect is only significant within the errors
for the skewness $p=3$ (greater than $2\sigma$ level). It is not clear whether 
this is driven by a pure luminosity dependence of the higher
order clustering or by a change in the galaxy mix with luminosity,
with different galaxy types having different $S_p$
or by a combination of the two effects: see
Norberg et al. (2002a) for an investigation of this point for the
2-point correlation function.
A simple linear relative bias model (dotted line in Fig.~10) 
does not reproduce the dependence of the $S_p$ on luminosity.

\end{itemize}

We have interpreted our results in terms of a simple, local bias
model, and we have quantified trends in clustering amplitude with
luminosity by estimating relative bias factors. These measurements,
summarised in Table~2, extend the constraints upon models of galaxy
formation derived from the two-point correlation function, quantifying
the shape of the tails of the count probability distribution as well
as its width. 

\section*{Acknowledgements}

The 2dFGRS was undertaken using the two-degree field spectrograph on
the Anglo-Australian Telescope. We acknowledge the efforts of all
those responsible for the smooth running of this facility during the
course of the survey, and also the indulgence of the time allocation
committees.  Many thanks go to Saleem Zaroubi and Simon White for
useful discussions and theoretical insight.  DC acknowledges the
financial support of the International Max Planck Research School in
Astrophysics Ph.D. fellowship, under which this work was carried out.
EG acknowledges support from the Spanish Ministerio de Ciencia y
Tecnologia, project AYA2002-00850 and EC FEDER funding.  CMB is
supported by a Royal Society University Research Fellowship.  PN
acknowledges the financial support through a Zwicky Fellowship at ETH,
Zurich.

\label{lastpage}

\end{document}